\documentclass[twocolumn,showpacs,prl,10pt,aps,floatfix]{revtex4-1}
\usepackage[english]{babel}
\usepackage{amsmath,amssymb}
\usepackage{times}
\usepackage{epsfig}
\usepackage[colorlinks,linkcolor=blue,citecolor=blue,urlcolor=blue]{hyperref}
\usepackage{color}
\begin{document}
\title{Spin and magnetothermal transport in the $S=1/2$ XXZ chain}
\author{C. Psaroudaki$^{1}$}
\author{X. Zotos$^{2,3,4}$}
\affiliation{$^1$ Department of Physics, University of Basel, 
Klingelbergstrasse 82, CH-4056 Basel, Switzerland}
\affiliation{$^2$Department of Physics, 
Crete Center for Quantum Complexity and Nanotechnology and\\ 
Institute of Theoretical and Computational Physics,
University of Crete, 71003 Heraklion, Greece}
\affiliation{$^3$Foundation for Research and Technology - Hellas, 71110 Heraklion, Greece}
\affiliation{$^4$ 
Max-Planck-Institut f\"ur Physik komplexer Systeme, 
N\"othnitzer Strasse 38, 01187 Dresden, Germany} 
\date{\today}

\begin{abstract}
We present a temperature and magnetic field dependence study of spin transport and magnetothermal corrections 
to the thermal conductivity in the spin $S=1/2$ integrable easy-plane regime Heisenberg chain, 
extending an earlier analysis based on the Bethe ansatz method. 
We critically discuss the low temperature, weak magnetic field behavior, the effect of magnetothermal 
corrections in the vicinity of the critical field and their role in recent thermal conductivity 
experiments in 1D quantum magnets.
\end{abstract}
\maketitle


Thermal transport by magnetic excitations is a research domain of actual interest 
where theoretical concepts are confronted and converge with state of the art experiments. The synthesis of high quality 
quasi-one dimensional quantum magnets allows the study of magnetic thermal conduction in spin liquids states, gapped and 
exotic topological excitation systems \cite{SpinMaterials}.
It is also amusing that prototype models used in the description of these systems, 
as the $S=1/2$ Heisenberg model, turn out to be totally unconventional, exhibiting ballistic transport at all temperatures 
due to the underlying integrability of the model \cite{Zotos97}. 

So far most thermal conductivity experiments are done on materials as the Sr$_2$CuO$_3$, SrCuO$_2$ or the ladder 
Sr$_{14}$Cu$_{24}$O$_{41}$ cuprate compounds, where the magnetic exchange constant $J$ is of the order of 2000 K 
and thus a magnetic field is not expected to play a significant role.
Only a few experiments in low $J$ (of the order of 10 K) compounds exist \cite{Sologubenko,Sun,Kohama} 
that pose the problem of magnetothermal corrections in thermal transport. 

In experiments, the measured thermal conductivity includes contributions from all itinerant particles 
or quasi--particles, such as charge carriers, spin excitations, phonons. In the case of insulators 
the study of thermal transport as a function of magnetic field is particularly 
attractive, as the magnetic field provides a handle to separate the field--independent 
phononic contribution from the total measured thermal conductivity \cite{Sologubenko}, confronting subtle
theoretical analysis to experiments. 
Furthermore, several intriguing phenomena in which the interplay of spin and heat transport play a crucial role have been 
suggested \cite{Louis,sakai2005,Furukawa,adachi}. In analogy to the thermoelectric effect in electronic conductors 
a spin - Seebeck effect should arise in the presence of a temperature gradient in electronic insulators.
Over the last few years, a great deal of experimental work has demonstrated such a generation of spin currents in 
a variety of (anti)ferro-magnetic insulating materials \cite{uchida} attracting interest to the novel field 
of {\it spin-caloritronics} \cite{bauer}.

In addition, one important aspect of the study of thermal transport as a function of magnetic field is the behavior 
of the various transport quantities close to the critical field $H_{cr}$, that corresponds to a 
Quantum Critical Point (QCP). The presence of a QCP can significantly affect the thermodynamic properties 
of a quantum magnet, such as magnetization or specific heat. However, new insights on the QCPs could be provided 
by the thermal and spin transport, and consequently by the thermomagnetic coefficients.

Within linear response theory the spin and energy current operators are defined from the continuity equation 
for the density of the local spin component $S_n^z$ and local energy correspondingly. 
For the Heisenberg chain, 
\begin{equation}
\mathcal{H}=\sum_{n=1}^{N} J(S_{n}^xS_{n+1}^x+S_{n}^yS_{n+1}^y+\Delta~S_{n}^zS_{n+1}^z)+H~S_{n}^z \,,
\label{Hamiltonian}
\end{equation}

\noindent
where $S_{i}^{\alpha}=\frac{\sigma_{i}^{\alpha}}{2}$ are the Pauli spin operators 
with components $\alpha=\lbrace x,y,z \rbrace$. The continuity equations lead to 
the spin $\mathcal{J}_s=J \sum_{n}(S_{n}^xS_{n+1}^y-S_n^yS_{n+1}^x)$,   
energy $\mathcal{J}_E=J^2\sum_{n} \mathbf{S}_{n} \cdot (\mathbf{S}_{n-1} \times \mathbf{S}'_{n+1})$ 
($\mathbf{S}'_n=(S_n^x,S_n^y,\Delta S_n^z)$)  and heat $\mathcal{J}_{Q}=\mathcal{J}_E+H\mathcal{J}_s$
current operators\cite{Zotos97,Mahan}. $\mathcal{J}_Q$ and $\mathcal{J}_S$ are related to the gradients of magnetic 
field $\nabla H$ and temperature $\nabla T$ by the transport coefficients $C_{ij}$ \cite{Mahan} :
\begin{equation}
\begin{pmatrix} \mathcal{J}_Q \\ \mathcal{J}_s \end{pmatrix} =
\begin{pmatrix} C_{QQ} & C_{Qs} \\ C_{sQ} & C_{ss} \end{pmatrix}
\begin{pmatrix} -\nabla T \\ \nabla H \end{pmatrix}\,,
\label{MatrixEquation}
\end{equation}
where $C_{QQ}=\kappa_{QQ}$ ($C_{ss}=\sigma_{ss}$) is the heat (spin) conductivity. The coefficients $C_{ij}$ correspond 
to time--dependent current--current correlation functions and 
it is straightforward to see that due to Onsager's relations \cite{Mahan}, 
$C_{sQ}=\beta C_{Qs}$. The real part of $C_{ij}(\omega)$ can be decomposed into a $\delta$ 
function at $\omega=0$ and a regular part:
\begin{equation}
\mbox{Re}(C_{ij}(\omega))=2\pi D_{ij}\delta(\omega)+C^{\text{reg}}_{ij}(\omega)\,.
\label{RealPart}
\end{equation}

Unconventional ballistic behavior in the sense of non decaying currents is signalled by a finite Drude weight $D_{QQ,ss}$ 
implying a divergent conductivity. The integrability of a model characterized by the existence of nontrivial local 
conservation laws 
is directly related to the existence of finite Drude weights at all temperatures \cite{Zotos97}. 
To start with, it is well established that the energy current operator $\mathcal{J}_{E}$ of the $S=1/2$ \textit{XXZ} model 
coincides with the first nontrivial conserved 
quantity \cite{Huber}, the currents do not decay and the long time asymptotic of the energy current--current dynamic 
correlations is finite, implying a finite $D_{EE}$ at any temperature which has been evaluated using Bethe ansatz 
techniques \cite{Klumper02}. 

Concerning the spin transport the situation is more involved as the spin current does 
not commute with the Hamiltonian. Nevertheless, it was shown \cite{Zotos97} using an inequality proposed 
by Mazur and Suzuki \cite{Mazur} that for several quantum integrable systems $D_{ss}$ is bounded 
by the thermodynamic overlap of the current operator with at least one conserved quantity. 
Unfortunately for the Hamiltonian 
of the $S=1/2$ model, all local conservation laws are invariant under spin inversion, 
whereas the spin current operator $\mathcal{J}_s$ is odd giving no useful bound at zero magnetic field. 
The existence of a finite $D_{ss}$ at finite $T$, as found by a BA approach \cite{Zotos99,Benz} 
has proven to be a delicate theoretical question for the zero magnetic field case. 
\begin{figure}[!t]
\includegraphics[width=0.8\columnwidth]{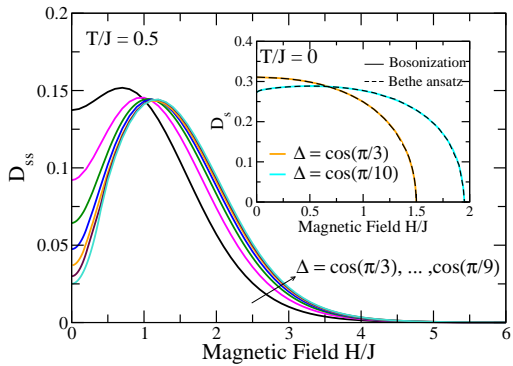}
\caption{(Color online) Magnetic field dependence of $D_{ss}$ at $T/J=0.5$ and several values 
of the anisotropy parameter $\Delta$. The inset depicts the magnetic field dependence of $D_{ss}$ at $T=0$ 
and two values of the 3nisotropy parameter $\Delta=\mbox{cos}(\pi/3),\mbox{cos}(\pi/10)$. Solid lines correspond to 
results obtained from bosonization and dashed lines from Bethe ansatz.}
\label{Ds_T05}
\end{figure}
Not until recently an improved Mazur bound was obtained \cite{ProsenOC} using a different approach based 
on deriving a whole family of almost conserved quasilocal conservation laws for an open XXZ chain up to boundary terms. 
It turns out that the quasilocal operator, with different symmetry properties than the local ones, 
has a finite overlap with $\mathcal{J}_s$ providing a nonzero lower bound for the spin Drude weight. 
This important result was later extended to the XXZ chain with periodic boundary conditions, where a family of exactly 
conserved quasilocal operators was constructed \cite{Pereira,ProsenPC}.

In the scope of these recent advances, we adress in this Letter the calculation of the spin Drude weight $D_{ss}$ 
in the presence of magnetic field. The calculation relies on a generalization of the approach 
proposed in Ref.~\cite{Zotos99} at zero magnetic field. The presence of a magnetic field results in some changes 
to the Bethe ansatz equations \cite{Takahashi}, but the overall analysis is essentially the same. 
The knowledge of $D_{ss}(T,H)$ also allows for the calculation of the thermal Drude weight $K_{th}$ and intriguing 
magnetothermal phenomena that arise due to the coupling of the energy and spin currents \cite{Louis}. Theoretically the problem of transport in the Heisenberg $S=1/2$ chain has been adressed by mean--field methods plus relaxation time approximation \cite{meisner05} and a combination of numerical exact diagonalization as well as Bethe ansatz techniques \cite{sakai2005,Furukawa}. 

A certain simplification of the Bethe ansatz equations for the massless regime $0 \leq \Delta \leq 1$  is provided 
under the parametrization $\Delta=\mbox{cos}(\pi/\nu)$, with integer $\nu$.  
The main results of this approach are that in the gapless regime $0 \leq \Delta \leq 1$, $D_{ss}(T,H=0)$ is nonzero 
with power--law behavior at low temperatures as:
\begin{equation}
D_{ss}(T,H=0)-D_{ss}(0,H=0)\sim - T^{\alpha}, 
\label{powerlaw}
\end{equation}

\noindent
$\alpha=\frac{2}{\nu-1}$, while in the high temperature limit $\beta \rightarrow 0$ the spin Drude weight behaves like 
$D_{ss}(T,H=0) = \beta C(\Delta)$ \cite{Benz}, where $C(\Delta)$ equals:
\begin{equation}
C(\Delta)= \frac{1}{16} \Big( 1-\frac{\sin(\frac{2\pi}{\nu})}{\frac{2\pi}{\nu}} \Big) \,,
\label{DshighT}
\end{equation}

\noindent
a result interestingly coinciding with the improved lower bound \cite{ProsenOC} at $\Delta=\cos(\pi/\nu)$.

At zero temperature the calculation of the magnetic field dependence of the spin Drude weight is feasible by considering the low--energy effective Hamiltonian of the \textit{XXZ} model using abelian bosonization. 
Within the Luttinger Liquid description, the spin Drude weight is expressed as $ D_{ss}=u(\Delta,H)K(\Delta,H)$,
\begin{figure}[!t]
\includegraphics[width=0.8\columnwidth]{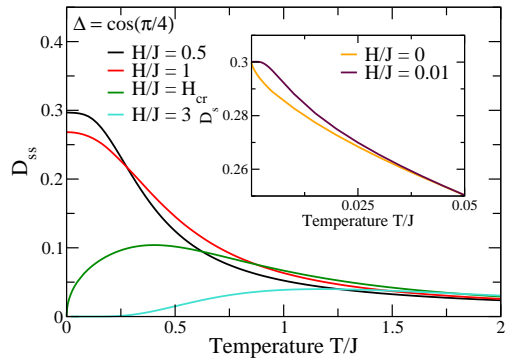}
\caption{(Color online) Temperature dependence of $D_{ss}$ for $\Delta=\mbox{cos}(\pi/4)$ and various magnetic fields. 
The inset depicts the $H=0$ power--law behavior of $D_{ss}$ at low temperatures given by Eq.\eqref{Ds_LowT}. The presence 
of small magnetic field $H/J=0.01$ suffices to destroy this singular behavior. }
\label{Ds_Pi4}
\end{figure}
where the Fermi velocity $u(\Delta,H)$ and the so-called Luttinger parameter $K(\Delta,H)$ depend on both the magnetic 
field $H$ and anisotropy parameter $\Delta$. For $ H=0$ they can be found in closed form \cite{Shastry90},
while at finite magnetic field, both parameters can be computed exactly from the Bethe ansatz 
solution \cite{BogoliubovBook}. 

We now turn our attention to the magnetic field dependence of $D_{ss}$ at finite temperature. 
In Fig.~\ref{Ds_T05} we depict $D_{ss}$ as a function of magnetic field $H$ for $T/J=0.5$ and various values 
of the anisotropy $\Delta$. The inset depicts the $D_{ss}(H)$ curve at $T=0$, calculated using the Luttinger Liquid 
description 
and the Bethe ansatz technique. The lines are indistinguishable providing a test of the Bethe ansatz calculation. 
We also find, as expected, that $D_{ss}(H)$ vanishes for $H > H_{cr}=J(1+\Delta)$, as the system enters 
its massive phase.  
The facts that become apparent from Fig.~\ref{Ds_T05} are the following: (i) At small magnetic fields the spin 
Drude weight goes like $D_{ss}(T,H)-D_{ss}(T,0) \simeq A H^2$, a behavior that is significantly different 
from the one at $T=0$. 
(ii) Upon increasing the magnetic field, $D_{ss}$ increases until it reaches a maximum and then it exponentially 
goes to zero. 
In the vicinity of $H_{cr}$, $D_{ss}$ is a smooth function of $H$ that is in direct contrast with the $T=0$ result. 
(iii) Upon increasing $\Delta$, starting from $\Delta=1/2$ and approaching the isotropic point $\Delta=1$, 
and for magnetic fields $H/J\gtrsim 0.5$, $D_{ss}$ seems to converge to a limiting behavior. 
 This is not true for small magnetic fields $H/J \lesssim 0.5$, where such a convergence should not be expected. 
The $D_{ss}(H=0)$ value strongly depends on $\Delta$ and goes to zero as $\Delta \rightarrow 1$ \cite{Zotos99}. 
\begin{figure}[!t]
\includegraphics[width=0.8\columnwidth]{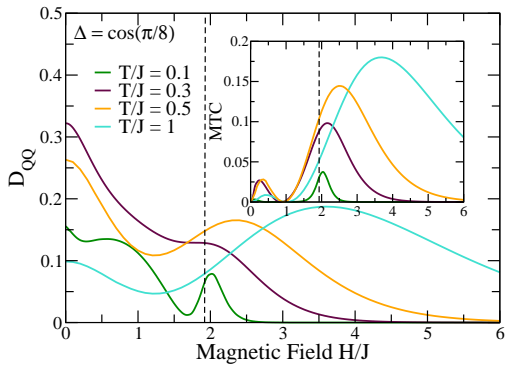}
\caption{(Color online) Magnetic field dependence of heat Drude weight $D_{QQ}$ at $\Delta=\cos(\pi/8)$ and several values 
of temperature $T$. The inset depicts the magnetic field dependence of $MTC$ term at the same $\Delta$. 
Vertical dotted line denotes position of $H_{cr}$.}
\label{Dqq_T}
\end{figure}

The temperature dependence of the spin Drude weight is also studied for four typical magnetic fields 
at $\Delta=\mbox{cos}(\pi/4)$ and the main features are depicted in Fig.~\ref{Ds_Pi4}. 
At $H\ll J$ the system is at its gapless phase, $D_{ss}$ is finite and at small temperatures it decreases like:
\begin{equation}
D_{ss}(T,H) - D_{ss}(T,0) \sim - e^{-H/T}T^{\gamma(H,\Delta)} \,,
\label{Ds_LowT}
\end{equation}

\noindent
where the exponent $\gamma$ depends on both $H$ and $\Delta$. At elevated temperatures, 
the $D_{ss}(T)$ curve vanishes as $1/T$. As shown in the inset of Fig.~\ref{Ds_Pi4}, 
the low $T$ behavior is in contrast with the $H=0$ results \cite{Zotos99} as the power--law of Eq.\eqref{powerlaw}, 
attributed to enhanced half-filling Umklapp scattering, is attenuated at $T < H$ . 
At $H=H_{cr}$ the system enters its gapped regime and $D_{ss}$ vanishes at $T=0$. Nevertheless, it becomes 
finite upon a small increase of temperature, exhibiting a $\sqrt{T}$ critical behavior at low $T$. 
The curve increases with $T$ until it reaches a maximum and then drops as $1/T$. 
Finally, in the gapped $H>H_{cr}$ regime we notice that at low $T$ the Drude weight is exponentially 
activated upon increase of $T$ and vanishes after a maximum. 
This behavior is summarized in Fig.~\ref{Ds_Pi4}. Also note that in the high temperature limit, 
the spin Drude weight behaves 
like $D_{ss}(T )=\beta C(\Delta)$, where $C(\Delta)$ is given by Eq.\eqref{DshighT}. 

Interesting conclusions can be drawn by a comparison to the special case 
of $\Delta=0$ (\textit{XY}) model, where both the spin and thermal currents 
are conserved and exact results can be found using the Jordan--Wigner 
transformation \cite{Furukawa}. The majority of features of $D_{ss}$ 
presented here for the $0<\Delta<1$ model are also realized for the $\Delta=0$ 
case. Nonetheless, one should emphasize that the power--law behavior 
of Eq.\eqref{powerlaw} is absent for the \textit{XY} model.

\begin{figure}[!t]
\includegraphics[width=0.8\columnwidth]{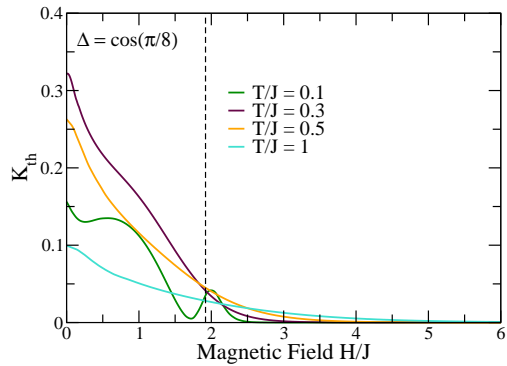}
\caption{(Color online) Magnetic field dependence of thermal Drude weight $K_{th}$ at $\Delta =\cos(\pi/8)$ 
and several values of temperature $T$.
Vertical dotted line denotes position of $H_{cr}$.}
\label{Kth_T}
\end{figure}
Now, with the novel input of $D_{ss}$ from the Bethe Ansatz analysis above, we can address the 
evaluation of thermal conductivity and magnetothermal coefficients.
To relate correlation functions $C_{ij}$ to experimentally accessible quantities 
we note that the spin conductivity $\sigma$ measured under the condition of $\nabla T=0$ is equal 
to $\sigma(\omega)=C_{SS}(\omega)$ and the thermal conductivity under the assumption of vanishing spin 
current $\mathcal{J}_s=0$, which is relevant to certain experimental setups, is redefined as follows:
\begin{equation}
\kappa(\omega)=C_{QQ}(\omega)-\beta \frac{C_{Qs}^2(\omega)}{C_{ss}(\omega)}\,,
\label{kmagc}
\end{equation}
where the second term is usually called the magnetothermal correction. Such a term originates from the coupling 
of the heat and spin currents in the presence of magnetic field \cite{Louis,sakai2005,meisner05} 
and is absent when $H=0$. In the case of ballistic transport, the thermal conductivity $K_{th}$ is found by combining 
Eqs.\eqref{kmagc} and \eqref{RealPart}: 
\begin{equation}
K_{th}=D_{QQ}-\beta \frac{D_{Qs}^2}{D_{ss}}\,.
\end{equation}
The first term $D_{QQ}$ corresponds to the heat conductivity, 
while the second term is the magnetothermal correction $MTC=\beta \frac{D_{Qs}^2}{D_{ss}}$. 
We should stress, in view of experiments \cite{Sologubenko,Sun,Kohama}, that this relation holds 
only when we assume the same relaxation rates for the magnetization and energy transport, an assumption deserving 
further study as it is not generally valid when inelastic processes are present.

It becomes apparent that $D_{QQ}$ and $K_{th}$ are the main quantities which play a central role in the study of 
thermal conductivity in the $S=1/2$ \textit{XXZ} chain. 
The thermal Drude weight $K_{th}$ is the result of a combination of two competing terms, the $D_{QQ}$ and $MTC$ term 
and for a complete picture of the thermal transport of the model all three terms need to be explored.
One can decompose the heat Drude weight $D_{QQ}$ in terms of the energy and spin contribution, which yields:
\begin{equation}
D_{QQ}=D_{EE}+2\beta H D_{Es}+\beta H^2 D_{ss} \,.
\label{Dqq_De}
\end{equation}
Similarly the $MTC$ term, and consequently the $K_{th}$ term, can be decomposed in terms of 
$D_{EE}$, $D_{Es}$ and $D_{ss}$. 
The $D_{EE}$ and $D_{Es}$ at finite temperatures have been calculated by Sakai and Kl\"{u}mper \cite{sakai2005} 
using a lattice path integral formulation, where a quantum transfer matrix (QTM) in imaginary time is introduced. 
This method produces all relevant correlations by solving two nonlinear integral equations at 
arbitrary magnetic fields and temperatures.

Let us begin by considering the magnetic field dependence of the various quantities. In Fig.~\ref{Dqq_T} we depict the 
heat Drude weight $D_{QQ}$ as a function of $H$ for various values of $T$ and $\Delta=\cos(\pi/8)$. 
An important fact of Fig.~\ref{Dqq_T} is that $D_{QQ}(H)$ exhibits a pronounced nonmonotonic behavior as a function of $H$. 
At small magnetic fields it decreases quadratically and then it rises again creating a peak before it vanishes at large 
magnetic fields. 

Next, we consider the behavior of the $MTC$ term as a function of $H$ as illustrated in the inset of Fig.~\ref{Dqq_T} 
for several $T$'s and $\Delta=\cos(\pi/8)$. 
As expected, the $MTC$ term is exactly zero at $H=0$, but becomes finite at finite $H$, where it develops two peaks 
with the second being more dominant than the first. 
The $MTC$ term turns out to be significant and should be taken into account for a complete description of thermal transport. 

The resulting behavior of the total thermal Drude weight $K_{th}$, as a sum of two competing terms, 
is summarised in Fig.~\ref{Kth_T}, where it is plotted as a function of $H$ for different temperatures. 
Fig.~\ref{Kth_T} allows 
for two major observations: (i) at $T/J \gtrsim 0.3$ the thermal Drude weight turns out to 
be a smooth function of magnetic field with 
no peaks observed as a function of $H$. The inclusion of the $MTC$ term results in an overall suppression of $K_{th}$ 
and the cancellation of the nonmonotonic peaked behavior of $D_{QQ}$. At higher temperatures the $MTC$ 
and $D_{QQ}$ terms develop a peak located exactly at the same field; the subtraction of these two terms results at 
a $K_{th}$ that is a smooth function of $H$. This finding is consistent with a numerical study of the thermal transport 
in the $S=1/2$ \textit{XXZ} chain in the presence of a magnetic field \cite{meisner05} based on exact diagonalization 
of a finite chain. 

Concerning thermal conductivity experiments \cite{Sologubenko,Sun,Kohama} in a magnetic field, although not conclusive, 
it seems that the rather featureless field dependence indicated in Fig.~\ref{Kth_T} (at $T/J \gtrsim 0.1$) 
is not observed but rather the nonmonotonic 
one shown in Fig.~\ref{Dqq_T}. The absence of MTC corrrections was attributed to the nonconservation of total 
magnetization due to spin-orbit scattering, but it could also be due to vastly different relaxation times 
for magnetization and energy transport. 

\begin{figure}[!t]
\includegraphics[width=0.8\columnwidth]{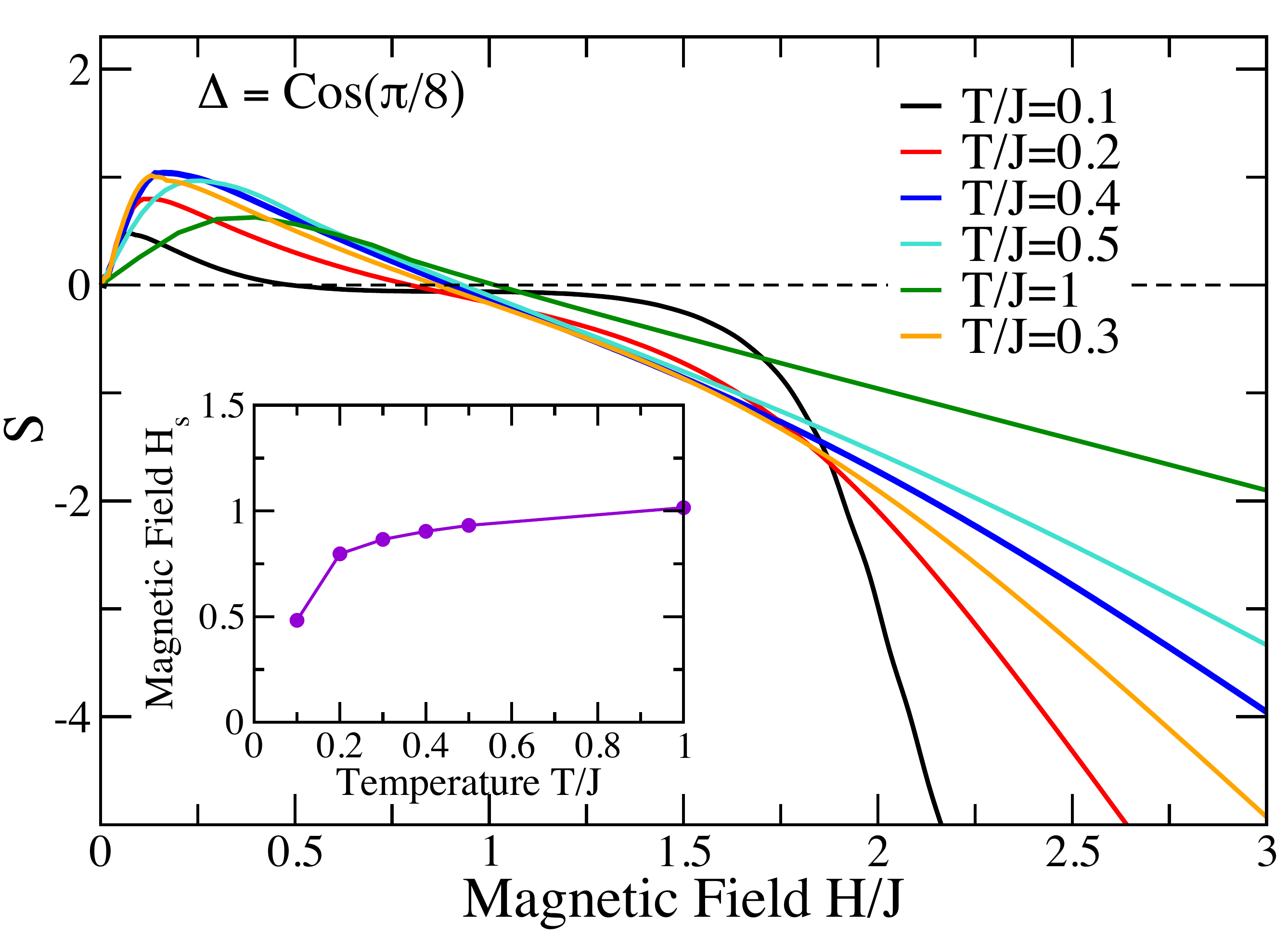}
\caption{(Color online)Thermal Seebeck coefficient $S$ for $\Delta=\mbox{cos}(\pi/8)$ and several values of $T$ as a function of $H$. The inset depicts the magnetic field $H_s$ at which $S$ changes sign, as a function of $T$. }
\label{SB}
\end{figure}

Finally, considering magnetothermal effects using Eq.\eqref{MatrixEquation}, the magnetic Seebeck coefficient $S$ 
under the condition of zero spin current $\mathcal{J}_s=0$ and for ballistic transport is given by,
\begin{equation}
S=\frac{\nabla H}{\nabla T}=\frac{C_{sQ}}{C_{ss}}=\frac{D_{sQ}}{D_{ss}} \,.
\end{equation}

\noindent
Here we take advantage of the Bethe ansatz technique to calculate $S$ as a function of $H$ for various temperatures 
in the thermodynamic limit. 
In Fig. \ref{SB} we depict the magnetic field dependence of $S$ for $\Delta=\mbox{cos}(\pi/8)$ and several values of $T$. 
We note that at small magnetic fields $S$ is positive, while at a certain magnetic field $H_s$ it changes sign and remains 
negative. In Ref.~\cite{Furukawa} it was suggested that the sign of $S$ is a criterion to clarify the types of carriers; 
a positive (negative) $S$ implies that the spin and heat are dominantly carried by carriers with up (down) spin. 
Upon increasing $T$ the structure of $S$ changes, but at any $T$ there is a single $H_s$ at which the Seebeck coefficient 
changes sign (see inset in Fig.~\ref{SB}). 

\acknowledgements{We would like to thank A. Kl\"{u}mper and J. Herbrych for stimulating discussions. This work was supported by the European Union (European Social Fund, ESF), Greek national funds through the Operational
Program ``Education and Lifelong Learning'' of the NSRF under ``Funding of proposals that have received
a positive evaluation in the 3rd and 4th call of ERC Grant Schemes'' and the European Union Program
No. FP7-REGPOT-2012-2013-1 under Grant No.~316165. }

%

\end{document}